\documentclass[12pt]{article}
\usepackage{amsmath}

\usepackage{amssymb}
\usepackage{bm}
\usepackage{braket}
\usepackage[dvips]{graphicx}

\begin{document}

\baselineskip=17pt

\begin{titlepage}
\rightline\today

\begin{center}
\vskip 2.5cm
\baselineskip=22pt
{\Large \bf {Nonperturbative definition of closed string theory}}\\
{\Large \bf {via open string field theory}}
\end{center}
\begin{center}
\vskip 1.0cm
{\large Yuji Okawa}
\vskip 1.0cm
{\it {Institute of Physics, The University of Tokyo}}\\
{\it {3-8-1 Komaba, Meguro-ku, Tokyo 153-8902, Japan}}\\
okawa@g.ecc.u-tokyo.ac.jp

\vskip 2.0cm

{\bf Abstract}
\end{center}

\noindent
In typical examples of the AdS/CFT correspondence,
the world-sheet theory with holes in the presence of D-branes
is assumed to be equivalent
in a low-energy limit
to a world-sheet theory without holes
for a different background such as $AdS_5 \times S^5$.
In the case of the bosonic string,
we claim
under the assumption of this equivalence
that open string field theory on $N$ coincident D-branes
can be used to provide
a nonperturbative definition of closed string theory
based on the fact that the $1/N$ expansion
of correlation functions of gauge-invariant operators
reproduces the world-sheet theory with holes
where the moduli space of Riemann surfaces is precisely covered.

\end{titlepage}

\section{Introduction}
\setcounter{equation}{0}

One of the most important problems in theoretical physics
is to formulate quantum gravity in a consistent manner.
While the quantization of general relativity
in the framework of quantum field theory
turned out to be difficult,
it was found that string theory consistently describes
on-shell scattering amplitudes involving gravitons.
However, string theory only provides a perturbative definition
of such on-shell scattering amplitudes
with respect to the string coupling constant.

One possible approach to a nonperturbative formulation of string theory
would be to introduce a spacetime field for each oscillation mode of the string
and construct an action of those spacetime fields.
The resulting theory in this approach is called string field theory.
Since gravitons are described as states of the closed string,
a natural approach to quantum gravity
would be to construct closed string field theory.
For closed bosonic string field theory,
the gauge invariance of the classical action
turned out to be anomalous
and we need quantum corrections to the action at each loop order
to recover the gauge invariance~\cite{Zwiebach:1992ie}.
The existence of such closed string field theory is useful
when we handle phenomena such as vacuum shift and mass renormalization
in the perturbative string theory as reviewed in~\cite{deLacroix:2017lif}.
However, formulating closed string field theory
at the quantum level nonperturbatively
by the path integral does not seem to be promising
because of these quantum corrections to the action.
As the origin of the quantum corrections
is related to the decomposition of the moduli space of Riemann surfaces,
we do not expect any improvement of the situation
in the generalization to closed superstring field theory.

Then how can we formulate string theory nonperturbatively?
The typical origin of the string perturbation theory
is the $1/N$ expansion of gauge theories
with $N \times N$ matrix degrees of freedom~\cite{tHooft:1973alw}.
Following this remarkable insight of 't~Hooft,
the long history of research on string theory indicates
that string theory can be defined nonperturbatively
in terms of such gauge theory.

In this paper we claim that {\it open} string field theory
instead of {\it closed} string field theory
can play a role of such gauge theory
and can be used to provide a nonperturbative definition
of closed string theory.
In the rest of this paper we present five questions and their answers
which will lead us to this claim.
The first question we ask is
what kind of closed string theory we should consider.

\section{Five questions}
\setcounter{equation}{0}

\subsection{What kind of closed string theory should we consider?}
\label{subsection-2.1}

While the $1/N$ expansion in the gauge theories
has a structure of the genus expansion in string theory,
we do not see the smooth world-sheet picture
in Feynman diagrams of matrix fields written in the double-line notation.
One attempt to generate the smooth world-sheet picture
was to take the double scaling limit
of matrix models~\cite{Brezin:1990rb, Douglas:1989ve, Gross:1989vs}.
This successfully defines string theory nonperturbatively,
but it worked out only for low spacetime dimensions
where physical degrees of freedom of gravitons are absent.

Then the conjecture called the AdS/CFT correspondence~\cite{Maldacena:1997re}
was put forward,
and it can be regarded as providing a nonperturbative definition of closed string theory
in terms of a quantum field theory without containing gravity.
Type IIB superstring theory on $AdS_5 \times S^5$, for example,
is conjectured to be defined nonperturbatively
by $\mathcal{N}=4$ $U(N)$ super Yang-Mills theory in four dimensions,
and the string coupling constant of
type IIB superstring theory on $AdS_5 \times S^5$ is given by $1/N$
in accord with the idea by 't~Hooft.
To understand how closed string theory appears from a theory without gravity
in this conjecture,
let us recall the standard explanation of the AdS/CFT correspondence
following section~3.1 of the review~\cite{Aharony:1999ti}.

Consider type IIB superstring theory on a flat spacetime in ten dimensions
with $N$ coincident D3-branes.
In the low-energy region where the energy of the system
is much lower than the string scale $1/\sqrt{\alpha'}$,
closed strings and open strings are decoupled.
Then closed string theory becomes a free theory in ten dimensions
and open string theory becomes $\mathcal{N}=4$ $U(N)$ super Yang-Mills theory
in four dimensions.

Next consider type IIB superstring theory on
the three-brane solution of supergravity.
Because of the redshift factor,
an object brought closer and closer to the three-brane
appears to have lower and lower energy for the observer at infinity.
In the same low-energy limit,
excitations propagating in ten dimensions
and excitations in the near horizon region are decoupled,
and we have a free theory in ten dimensions
and type IIB superstring theory on $AdS_5 \times S^5$,
which is the near horizon geometry of the three-brane solution.
We are then led to the conjecture
that $\mathcal{N}=4$ $U(N)$ super Yang-Mills theory in four dimensions
is the same as type IIB superstring theory on $AdS_5 \times S^5$.

As can be seen from this explanation,
the AdS/CFT correspondence tells us
that the world-sheet theory with holes
in the presence of D-branes
is equivalent to a different world-sheet theory
on a curved background in the low-energy limit.
This equivalence was shown
in the context of the large $N$ duality of the topological string~\cite{Gopakumar:1998ki}
by Ooguri and Vafa~\cite{Ooguri:2002gx}.
For developments in the superstring,
see, for example, \cite{Berkovits:2003pq, Berkovits:2007rj, Berkovits:2019ulm}.
While establishing this equivalence is a crucial ingredient
for proving the AdS/CFT correspondence,
we assume this equivalence in this paper
and we instead concentrate on two other aspects.
The first aspect is to see that the world-sheet theory with holes
is a consistent perturbation theory
which can be interpreted as a theory of closed strings.
The second aspect is how the world-sheet theory with holes
can be reproduced by a theory from the open string sector.

Let us begin with the question
of whether the world-sheet theory with holes
can be interpreted as a consistent perturbation theory of closed strings.
Compared with the discussion in the topological string,
we have to be more careful in the physical string theory.
First of all, the moduli space of Riemann surfaces with holes
must be covered for consistency.
In the moduli space there are regions where two boundaries are close,
and such a region corresponds to propagation of an open string.
So we may think that open strings are necessary for unitarity,
and this should be a theory of closed strings and open strings.
We claim that this is not necessarily the case
in the context relevant for the AdS/CFT correspondence.
First, in this context we focus on a sector
which is analogous to the gauge-invariant sector of gauge theory.
When we consider correlation functions of gauge-invariant operators
in gauge theory,
we never see poles from fields which are not gauge invariant.
Second, correlation functions of one closed string vertex operator
and one open string vertex operator on the disk are generically nonvanishing
so that the closed string propagation and the open string propagation
are mixed when the interaction is turned on.
On-shell states in the interacting theory
need to be identified by diagonalizing such propagators
and the long propagation of an open string
does not necessarily generates an on-shell pole.\footnote{
The situation is analogous to the mass renormalization
of closed string theory reviewed in~\cite{deLacroix:2017lif}.
On-shell states in the interacting theory
are generically different from those in the free theory.
}
While the world-sheet contains holes,
such on-shell states can be regarded as closed string states
just as we regard Wilson loops as closed strings
in gauge theory.
Feynman diagrams of gauge theory in the double-line notation
are similar to the world-sheet with holes,
but one important difference is
that the world-sheet with holes
should be associated with Riemann surfaces,
and the covering of the moduli space
of Riemann surfaces is crucial for consistency.
Note that the number of holes is irrelevant
for the consistent world-sheet picture
as long as the moduli space of Riemann surfaces is covered.\footnote{
In the hexagon approach based on the integrability~\cite{Basso:2015zoa},
the world-sheet picture appears
even for the weak-coupling region,
and our viewpoint might provide an explanation of this feature.
}
This should be contrasted with the world-sheet picture
which appears in the double scaling limit of the matrix models.

When we regard the world-sheet theory with holes
as closed string theory, what corresponds to the string coupling constant?
Let us consider the theory on $N$ coincident D-branes,
and we organize the Feynman diagrams
in terms the 't~Hooft coupling constant as usual.
Then the coupling constant of closed string theory
is given by $1/N$.
Assuming the equivalence
of this world-sheet theory with holes
to a world-sheet theory without holes for a different background,
we expect that this theory
contains gravity in the low energy,
and this is the perturbation theory
that we want to reproduce by a theory without gravity.
Let us now present our answer to the first question.\\

\noindent
\underbar{Question 1}
\begin{quotation}
\noindent
What kind of closed string theory should we consider?
\end{quotation}
\noindent
\underbar{Answer}
\begin{quotation}
\noindent
We should consider the world-sheet theory with holes
which can be interpreted as a consistent perturbation theory of closed strings
and which we expect to contain gravity.
\end{quotation}

It would be difficult to see this world-sheet picture
directly in $\mathcal{N}=4$ super Yang-Mills theory
because it is the theory after taking the low-energy limit.
Before taking the low-energy limit,
the dynamics on the D-branes is described by {\it open string field theory},
and gauge invariance of open string field theory
is closely related to the world-sheet picture.
Now the second question we ask is
what quantities we should consider in open string field theory.

\subsection{What quantities should we consider in open string field theory?}

In the context of the AdS/CFT correspondence,
we consider correlation functions of gauge-invariant operators
in $\mathcal{N}=4$ super Yang-Mills theory.
In string field theory, it is in general difficult
to construct gauge-invariant operators.
It is in fact an important feature of string field theory
and it is part of the reason that the interacting string field theory
is believed to be unique up to field redefinition
given a free theory.
In open bosonic string field theory
with the cubic interaction in terms of the star product~\cite{Witten:1985cc},
however, there are a class of gauge-invariant operators
and we can define a gauge-invariant operator
for each on-shell closed string vertex operator~\cite{Hashimoto:2001sm, Gaiotto:2001ji}.
These gauge-invariant operators have been mainly discussed
in the context of the classical theory.
They were evaluated for a classical solution
to extract information on the boundary conformal field theory
corresponding to the classical solution~\cite{Ellwood:2008jh, Kudrna:2012re}.
In that context they were called
gauge-invariant observables, gauge-invariant overlaps,
Ellwood invariants, and so on. 
We call them gauge-invariant operators
as we consider them in the quantum context.

The action of open bosonic string field theory~\cite{Witten:1985cc} is given by
\begin{equation}
S = {}-\frac{1}{2} \, \langle \, \Psi, Q \Psi \, \rangle
-\frac{1}{3} \, \langle \, \Psi, \Psi \ast \Psi \, \rangle \,,
\end{equation}
where $\Psi$ is the open string field of ghost number $1$,
$Q$ is the BRST operator,
$\langle \, A, B \, \rangle$ is the BPZ product of $A$ and $B$,
and $A \ast B$ is the star product of $A$ and $B$.
Before we explain the definition of the gauge-invariant operators,
it will be useful to recall the definitions of the BPZ inner product
and the star product.\footnote{
More detailed explanations can be found in the review~\cite{Okawa:2012ica},
where only the basic knowledge of conformal field theory
in Chapter~2 of the textbook by Polchinski~\cite{Polchinski:1998rq} is assumed.
}

\begin{figure}[bth]
\centerline{\includegraphics[width=5cm]{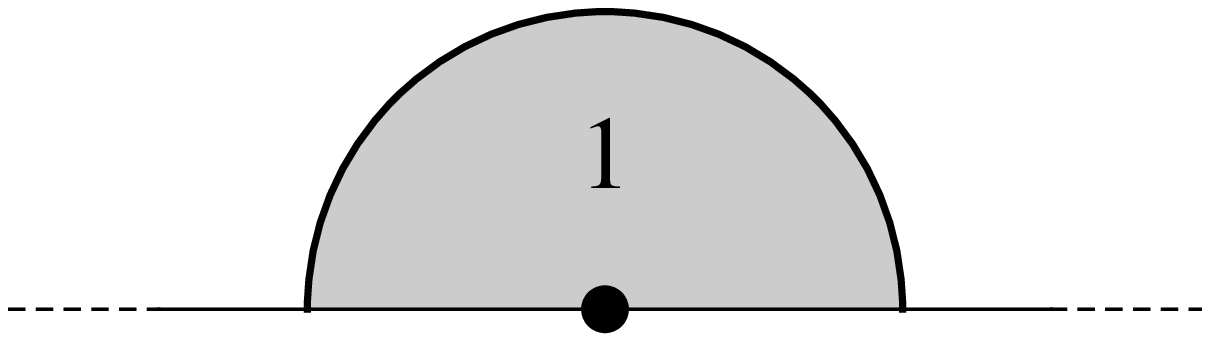}
\includegraphics[width=5cm]{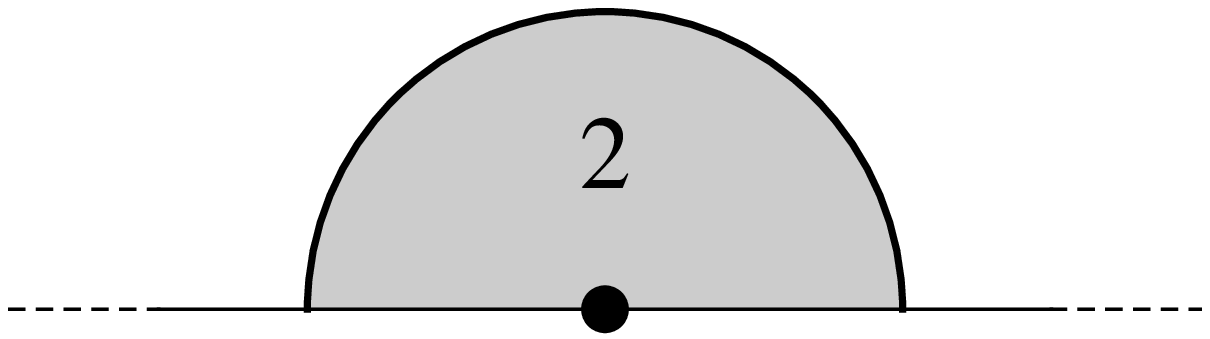}}
\vspace{0.5cm}
\centerline{$\downarrow$}
\vspace{0.5cm}
\centerline{\includegraphics[width=10cm]{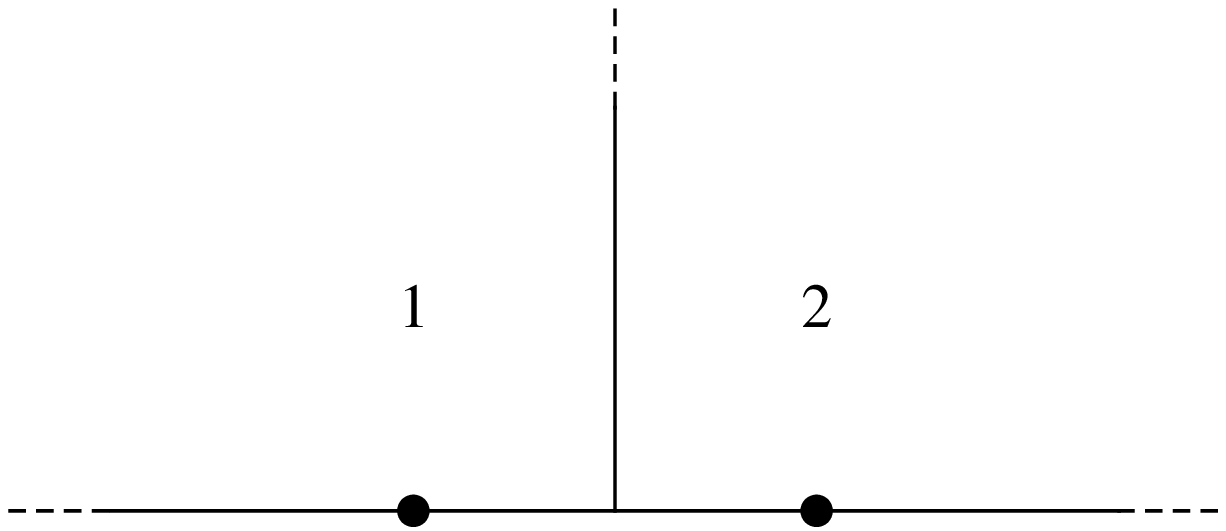}}
\caption{The definition of the BPZ inner product.}
\label{BPZ-inner-product-figure}
\end{figure}

The BPZ inner product for a pair of states $A_1$ and $A_2$ is defined
by the following two-point correlation function on the upper half-plane:
\begin{equation}
\langle \, A_1, A_2 \, \rangle
= \langle \, h_1 \circ A_1 (0) \, h_2 \circ A_2 (0) \, \rangle_{\rm UHP} \,,
\end{equation}
where $A_1 (0)$ and $A_2 (0)$ are the operators corresponding
to the states $A_1$ and $A_2$, respectively, in the state-operator correspondence.
Here and in what follows we denote the operator mapped from $\mathcal{O} (\xi)$
in the local coordinate $\xi$
under a conformal transformation $f(\xi)$ by $f \circ \mathcal{O} (\xi)$.
The conformal transformations $h_1 (\xi)$ and $h_2 (\xi)$ are given by
\begin{equation}
h_1 (\xi) = \tan \biggl( \arctan \xi -\frac{\pi}{4} \, \biggr)
= \frac{\xi-1}{\xi+1} \,, \qquad
h_2 (\xi) = \tan \biggl( \arctan \xi +\frac{\pi}{4} \, \biggr)
= {}-\frac{\xi+1}{\xi-1} \,.
\end{equation}
See figure~\ref{BPZ-inner-product-figure} for illustration of this definition.

\begin{figure}[bth]
\centerline{\includegraphics[width=4cm]{xi1.eps}
\includegraphics[width=4cm]{xi2.eps}
\includegraphics[width=4cm]{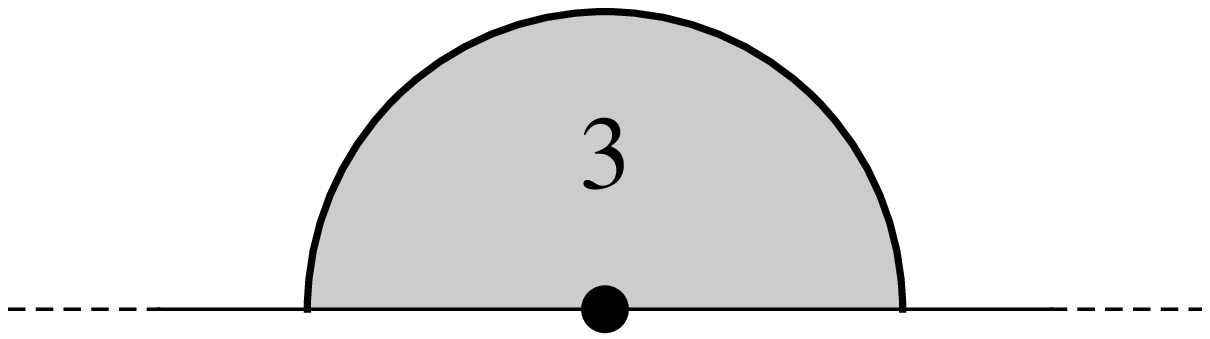}}
\vspace{0.5cm}
\centerline{$\downarrow$}
\vspace{0.5cm}
\centerline{\includegraphics[width=10cm]{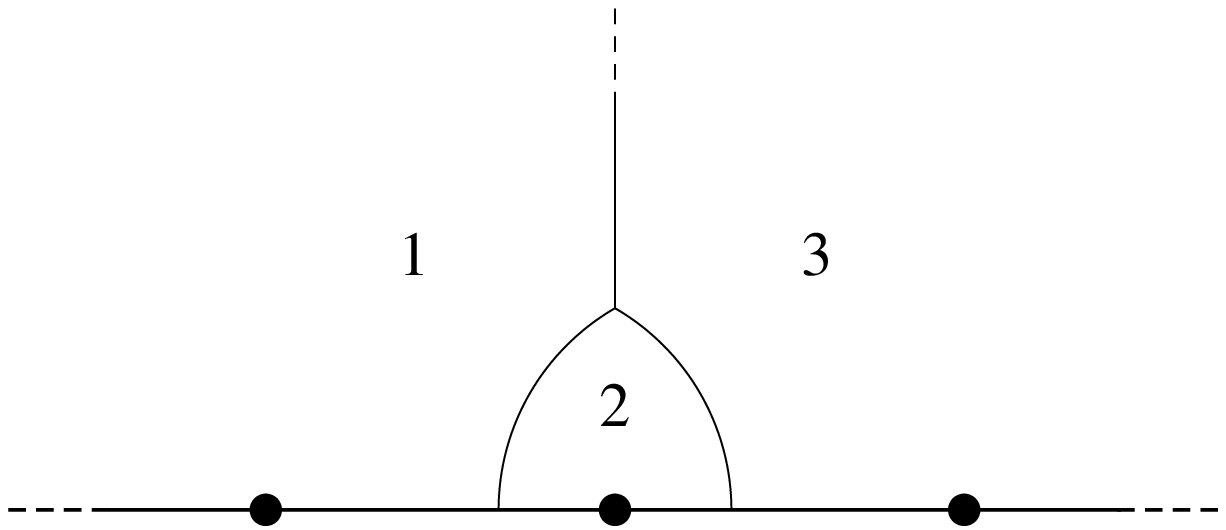}}
\caption{The definition of the star product.}
\label{star-product-figure}
\end{figure}

The star product is defined
by the following three-point correlation function on the upper half-plane:
\begin{equation}
\langle \, A_1, A_2 \ast A_3 \, \rangle
= \langle \, f_1 \circ A_1 (0) \, f_2 \circ A_2 (0) \, f_3 \circ A_3 (0) \, \rangle_{\rm UHP} \,,
\end{equation}
where $A_1 (0)$, $A_2 (0)$, and $A_3 (0)$ are the operators corresponding
to the states $A_2$, $A_2$, and $A_3$, respectively, in the state-operator correspondence
and the conformal transformations $f_1 (\xi)$, $f_2 (\xi)$, and $f_3 (\xi)$ are given by
\begin{equation}
\begin{split}
f_1 (\xi) & = \tan \biggl[ \, \frac{2}{3} \, \biggl(
\arctan \xi -\frac{\pi}{2} \, \biggr) \, \biggr] \,, \qquad 
f_2 (\xi) = \tan \biggl( \, \frac{2}{3} \,
\arctan \xi \, \biggr) \,, \\
f_3 (\xi) & = \tan \biggl[ \, \frac{2}{3} \, \biggl(
\arctan \xi +\frac{\pi}{2} \, \biggr) \, \biggr] \,. 
\end{split}
\end{equation}
See figure~\ref{star-product-figure} for illustration of this definition.

\begin{figure}[h]
\centerline{\includegraphics[width=5cm]{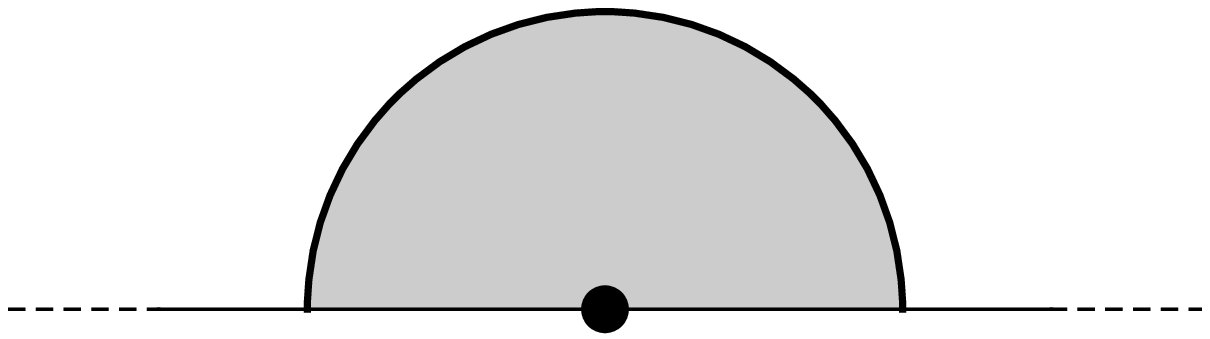}}
\vspace{0.5cm}
\centerline{$\downarrow$}
\vspace{0.5cm}
\centerline{\includegraphics[width=10cm]{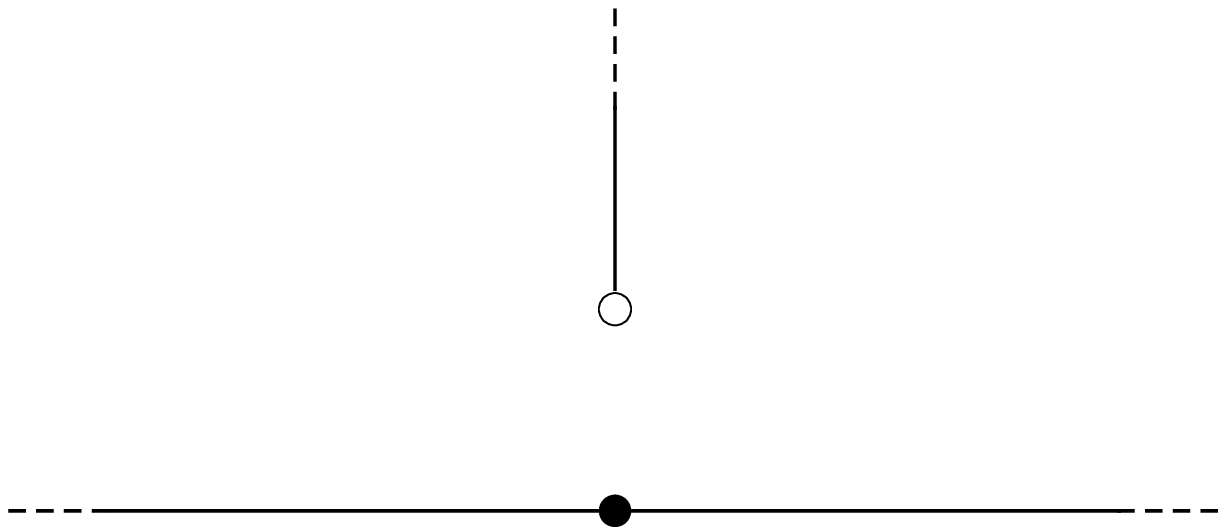}}
\caption{The definition of the gauge-invariant operator.
The interior of the half disk in the local coordinate $\xi$
is mapped to the whole upper half-plane
with the left half and the right half of the open string
being glued together.
The white dot represents the insertion of the vertex operator
for the on-shell closed string at the open-string midpoint.
}
\label{gauge-invariant-operator-figure}
\end{figure}

Let us finally explain the definition of the gauge-invariant operator.
The gauge-invariant operator $\mathcal{A}_\mathcal{V} [ \, \Psi \, ]$ 
for an on-shell closed string vertex operator $\mathcal{V}$ of ghost number $2$ is defined
by the following correlation function on the upper half-plane:
\begin{equation}
\mathcal{A}_\mathcal{V} [ \, \Psi \, ]
= \langle \, f_{\rm mid} \circ \mathcal{V} (0) \, f_I \circ \Psi (0) \, \rangle_{\rm UHP} \,,
\end{equation}
where $\Psi (0)$ is the operator corresponding
to the state $\Psi$ in the state-operator correspondence.
The conformal transformation $f_I (\xi)$
associated with the identity string field is given by
\begin{equation}
f_ I (\xi) = \tan \Bigl( \, 2 \, \arctan \xi \, \Bigr)
= \frac{2 \, \xi}{1-\xi^2} \,,
\end{equation}
and $f_{\rm mid} (\xi)$ is the translation to the open-string midpoint:
\begin{equation}
f_{\rm mid} (\xi) = \xi +i \,.
\end{equation}
See figure~\ref{gauge-invariant-operator-figure} for illustration of this definition.

These gauge-invariant operators have an interesting origin
in open-closed string field theory.
A one-parameter family of formulations
for open-closed bosonic string field theory
were constructed in~\cite{Zwiebach:1992bw},
and it was observed
that in a singular limit the action reduces
to that of the cubic open bosonic string field theory 
with an additional vertex
which couples one off-shell open string field
and one on-shell closed string field:
\begin{equation}
S = {}-\frac{1}{2} \, \langle \, \Psi, Q \Psi \, \rangle
-\frac{1}{3} \, \langle \, \Psi, \Psi \ast \Psi \, \rangle
+\langle \, J(\Phi) \,, \Psi \, \rangle \,,
\label{action}
\end{equation}
where $\Phi$ is the on-shell closed string field,
\begin{equation}
Q \Phi = 0 \,,
\end{equation}
and $J(B)$ is a map from a closed string field $B$
to an open string field.
The vertex $\langle \, J(B), A \, \rangle$
for an open string field $A$ and a closed string field $B$ 
is defined by
\begin{equation}
\langle \, J(B) \,, A \, \rangle
= \langle \, f_{\rm mid} \circ B (0) \, f_I \circ A (0) \, \rangle_{\rm UHP} \,,
\end{equation}
where $A (0)$ and $B (0)$ are the operators corresponding
to the states $A$ and $B$, respectively, in the state-operator correspondence.
The Grassmann parity of $J(B)$ is the same as that of $B$ mod $2$.
Two important properties associated with $J(\Phi)$ are as follows:
\begin{equation}
\begin{split}
Q J(\Phi) = 0 \,, \qquad
J(\Phi) \ast A = A \ast J(\Phi) \,,
\end{split}
\end{equation}
where $A$ is an arbitrary open string field.
The kinetic term of the closed string field is absent
so that the resulting theory is no longer
open-closed string field theory.
It is open string field theory,
and the coupling of the on-shell closed string field
and the off-shell open string field
can be regarded as source terms
for the gauge-invariant operators:
\begin{equation}
\langle \, J(\Phi) \,, \Psi \, \rangle
= \sum_\alpha \, \mathcal{G}_\alpha \, \mathcal{A}_{\mathcal{V}_\alpha} [ \, \Psi \, ] \,,
\end{equation}
where the collective label $\alpha$ generically contains
both continuous and discrete variables
and the summation over $\alpha$ should be understood
to include integrals for continuous variables.
The source $\mathcal{G}_\alpha$ for $\mathcal{A}_{\mathcal{V}_\alpha} [ \, \Psi \, ]$
is related to $\Phi$ via the expansion
\begin{equation}
\Phi = \sum_\alpha \, \mathcal{G}_\alpha \, \Phi_\alpha \,,
\end{equation}
where $\Phi_\alpha$ is the state corresponding to $\mathcal{V}_\alpha$
in the state-operator correspondence.

An important consequence from this relation
of the gauge-invariant operators and open-closed string field theory
is that Feynman diagrams for correlation functions
of the gauge-invariant operators
are given by Riemann surfaces containing holes
with bulk punctures
and the moduli space of such Riemann surfaces is covered.
Note that open bosonic string field theory
with the cubic vertex in terms of the star product
plays a distinguished role in this context.

Let us now consider the theory on $N$ coincident D-branes.
If we evaluate
correlation functions of the gauge-invariant operators
in the $1/N$ expansion,
by construction it reproduces
the perturbation theory we mentioned in~\S\ref{subsection-2.1}.
Let us present our answer to the second question.\\

\noindent
\underbar{Question 2}
\begin{quotation}
\noindent
What quantities should we consider in open string field theory?
\end{quotation}
\noindent
\underbar{Answer}
\begin{quotation}
\noindent
We should consider correlation functions of the gauge-invariant operators.
The moduli space of Riemann surfaces associated with Feynman diagrams
is covered, and the $1/N$ expansion
can be interpreted as a closed string perturbation theory.\\
\end{quotation}

In open string field theory
we can also consider dynamics of open strings
in addition to the gauge-invariant operators.
This may be related to the discussion
on non-singlet sectors in the matrix models~\cite{Maldacena:2005hi}.
While this is an interesting direction to explore,
we concentrate on correlation functions
of the gauge-invariant operators in this paper.

The action~\eqref{action} generates
the complete set of Riemann surfaces
containing an arbitrary number of holes
with an arbitrary number of bulk punctures
as Feynman diagrams for correlation functions
of the gauge-invariant operators,
but such Riemann surfaces contain at least one hole
and contributions from Riemann surfaces without any holes are missing.
In the context of on-shell scattering amplitudes,
they are necessary for factorization.
Then our third question is what we lose in the missing Feynman diagrams.

\subsection{What do we lose in the missing Feynman diagrams?}

Let us again recall the explanation of the AdS/CFT correspondence.
Both in the description with D-branes
and in the description with the three-brane solution of supergravity,
there are two decoupled sectors in the low-energy limit.
One of them
is a free theory in ten dimensions from the closed string sector,
and we identified the two descriptions of the other sector.
What we want is of course the interacting sector.
In the low-energy limit, the missing contributions
correspond to those of a free closed string theory,
which we wan to discard.
Therefore our answer to the third question is as follows.\\

\noindent
\underbar{Question 3}
\begin{quotation}
\noindent
What do we lose in the missing Feynman diagrams?
\end{quotation}
\noindent
\underbar{Answer}
\begin{quotation}
\noindent
Nothing in the low-energy limit!
\end{quotation}

Previously there were some attempts to reproduce
closed string theory without holes in the world-sheet from correlation functions
of the gauge-invariant operators, for example, using tachyon condensation.
While these attempts are interesting,
we emphasize that our approach is different
and we are not trying to reproduce closed string theory on a flat spacetime.

Note that after taking the low-energy limit
the quantities we are considering are no longer on-shell scattering amplitudes.
There may be a physical interpretation
about correlation functions of the gauge-invariant operators
before taking the low-energy limit,
but we have not figured it out.

In the low-energy limit, gauge-invariant operators
with vertex operators for massive closed string states
will not play an important role.
We emphasize that they are massive closed string states
in the presence of D-branes,
and massive closed string states in $AdS_5 \times S^5$
arise from the massless sector in the asymptotically flat spacetime
by taking the low-energy limit.
Under the assumption
of the equivalence between the world-sheet theory with holes
and a world-sheet theory without holes for a different background,
the $1/N$ expansion of correlation functions
of the gauge-invariant operators in the low-energy limit
should incorporate interactions
in terms of massive closed string states in the different background
such as $AdS_5 \times S^5$.

To summarize, we claim that the evaluation of correlation functions
of the gauge-invariant operators in the $1/N$ expansion
can be interpreted as a closed string perturbation theory
in the low-energy limit.
Therefore, 
if open string field theory for finite $N$ is a consistent quantum theory,
it provides a nonperturbative definition of closed string theory.
Now the fourth question we ask is
whether open string field theory is a consistent quantum theory.

\subsection{Is open string field theory a consistent quantum theory?}

In general, we do not expect open bosonic string field theory
to be a consistent quantum theory
because of the presence of tachyons
in the open string channel and in the closed string channel.
In the topological string or in the noncritical string,
however, there are backgrounds without tachyons
and it would be interesting to consider
the quantum theory for gauge-invariant operators
of open bosonic string field theory.
For example, three-dimensional Chern-Simons gauge theory can be formulated
as open string field theory on topological A-branes~\cite{Witten:1992fb}.
The duality in the B-model topological string theory
is also discussed recently~\cite{Costello:2018zrm},
and it would be interesting to consider open string field theory
in this context.
In the case of the noncritical string,
it was shown by Gaiotto and Rastelli
that the Kontsevich model~\cite{Kontsevich:1992ti}
can be realized as open string field theory~\cite{Gaiotto:2003yb}.
Another interesting arena is the recent discussion
on D-instanton contributions in two-dimensional
string theory~\cite{Balthazar:2019rnh, Sen:2019qqg, Balthazar:2019ypi, Sen:2020cef, Sen:2020oqr}.

On the other hand,
open superstring field theory can be a consistent quantum theory.
When we quantize open superstring field theory,
we know that both the Neveu-Schwarz sector
and the Ramond sector are necessary for consistency.
While the action of open superstring field theory
involving the Ramond sector had not been constructed for many years,
this problem was recently overcome
and we now have several formulations of open superstring field theory
which are complete
at the classical level~\cite{Kunitomo:2015usa, Sen:2015uaa, Erler:2016ybs, Konopka:2016grr}.
We consider that the formulations of open superstring field theory
need to be developed further
and it is an important question to address
whether or not open superstring field theory is consistent
as a quantum theory,
but at the same time we consider
that we are in a position
to discuss how we use open superstring field theory
to understand the mechanism which realizes
the AdS/CFT correspondence.
Let us now present our answer to the fourth question.\\

\noindent
\underbar{Question 4}
\begin{quotation}
\noindent
Is open string field theory a consistent quantum theory?
\end{quotation}
\noindent
\underbar{Answer}
\begin{quotation}
\noindent
Open bosonic string field theory for the topological string
or the noncritical string can be a consistent quantum theory.
Open superstring field theory
can also be a consistent quantum theory,
which motivates us to extend our discussion
to the superstring.
\end{quotation}

In our perspective gauge invariance of open string field theory
at the quantum level is crucially important.
For open bosonic string field theory,
the cubic theory in terms of the star product
is just one gauge-invariant formulation
and there are other formulations
which are also gauge invariant at the classical level.
At the quantum level, however, the cubic theory in terms of the star product
seems to play a distinguished role.
For open superstring field theory, as we mentioned before,
there are several formulations at the classical level,
but there might be a distinguished theory at the quantum level.

At any rate, we should seriously think about quantization
of open string field theory.
Our fifth question is then
whether we can make sense of the path integral of open string fields.

\subsection{Can we make sense of the path integral of open string fields?}

As we mentioned in the introduction,
we do not consider
the path integral of closed string field theory
to be promising for a nonperturbative definition of closed string theory.
On the other hand,
there will be a better chance
of making sense of the path integral of open string field theory.
However, it may be still difficult
because the open string field contains infinite component fields.

Actually, we define closed string theory
by taking the low-energy limit of open string field theory
so that we can in principle integrate out massive fields
of open string field theory
following the approach developed by Sen~\cite{Sen:2016qap}.
In general, the resulting theory in terms of massless fields
will be very complicated.
If we can identify a theory in the same universality class,
however, we can use it to define closed string theory nonperturbatively.

In the case of D3-branes, for example, the resulting theory
will be equivalent to $\mathcal{N}=4$ super Yang-Mills theory
in the low-energy limit.
As we mentioned before, it is difficult
to see the world-sheet picture
in $\mathcal{N}=4$ super Yang-Mills theory,
but we can keep track of the relation
to the world-sheet
in the theory after integrating out massive fields.
This can be a promising way
for proving the AdS/CFT correspondence.
Our answer to the fifth question is as follows.\\

\noindent
\underbar{Question 5}
\begin{quotation}
\noindent
Can we make sense of the path integral of open string fields?
\end{quotation}
\noindent
\underbar{Answer}
\begin{quotation}
\noindent
While there is a possibility of
making sense of the path integral of open string fields,
we can also consider
integrating out massive fields
to obtain a theory in terms of massless fields.
If we can identify a theory in the same universality class,
we can use it to define closed string theory nonperturbatively.
\end{quotation}

One puzzling feature in our approach is
that the gauge-invariant operator is a linear functional of the open string field
and apparently it does not look like operators which couple to closed strings
such as the energy-momentum tensor.
The resolution of this puzzle
is also related to taking the low-energy limit.
In the process of integrating out massive fields,
couplings of the closed string and multiple open string fields are generated,
and the gauge-invariant operators in terms of massless fields
will resemble single-trace operators of $U(N)$ gauge theories
in the low-energy limit~\cite{KOS, EMSV}.
We expect that the world-sheet calculations
of the energy-momentum tensor
of noncommutative gauge theory in~\cite{Okawa:2000sh}
or of the BFSS matrix model~\cite{Banks:1996vh} in~\cite{Okawa:2001if}
are reproduced by taking the low-energy limit in our approach.

We believe that there are several advantages in our approach.
First, we define correlation functions of the gauge-invariant operators
before taking the low-energy limit,
and this can provide a well-defined setting to discuss
the correspondence after taking the limit
such as the AdS/CFT dictionary between correlation functions on the boundary
and supergravity calculations in the bulk~\cite{Gubser:1998bc, Witten:1998qj}.
Second, we have a relation between
gauge-invariant operators and closed string states from the beginning,
and this can be an advantage,
although it might be difficult to keep track of the relation
after taking the low-energy limit.
Third, our discussion can be applied to any background
by taking an appropriate limit
discussed in~\cite{Douglas:1996yp, Sen:1997we, Seiberg:1997ad}.
In particular, our discussion does not directly rely
on conformal symmetry in the limit or on supersymmetry.

\section{Discussion}
\setcounter{equation}{0}

We want to have a consistent formulation of quantum gravity.
For this purpose
we want to define closed string theory nonperturbatively.
Instead of the ordinary world-sheet theory of the closed string,
we consider the world-sheet theory with holes in this paper.
Our expectation that this perturbation theory contains gravity
is based on the assumption that this theory is equivalent
to a world-sheet theory without holes for a different background.
This is a crucial assumption, and one approach to the proof
is to consider particular examples.
The most promising example would be the world-sheet theory
with holes where the boundary conditions for D3-branes are imposed,
and this theory is believed to be equivalent
to the world-sheet theory for the $AdS_5 \times S^5$ background
in the low-energy limit.
Another possible approach would be to use open-closed string field theory.
We can formally integrate out the open string field to show
that the resulting theory has the same algebraic structure
as closed string field theory,
although it is subtle to integrate out massless fields from the open string field.
Furthermore, the algebraic structure of closed string field theory
strongly indicates that it contains gravity,
but we do not know what background the resulting theory describes in general,
and we do not even know whether the background has a geometric interpretation.

The action~\eqref{action} for $N$ coincident D-branes generates
Feynman diagrams for the world-sheet theory with holes
where the moduli space of Riemann surfaces is precisely covered,
and this implies that the $1/N$ expansion of correlation functions
of the gauge-invariant operators reproduces
the perturbation theory which we expect to contain gravity.
This is the main observation of this paper.
The proof of the covering of the moduli space in~\cite{Zwiebach:1992bw}
is truly remarkable, and the challenge is to extend
the proof to the supermoduli space of super-Riemann surfaces.
Such an extension would establish
that open superstring field theory is a consistent quantum theory,
and we hope that such open superstring field theory
can be used to provide a nonperturbative formulation of quantum gravity
based on the scenario described in this paper.

\bigskip
\noindent
{\bf \large Acknowledgments}

\medskip
I would like to thank Shota Komatsu, Hirosi Ooguri, Ashoke Sen,
Yuji Tachikawa and Tamiaki Yoneya for useful discussions.
I also thank the Galileo Galilei Institute for Theoretical Physics
and INFN for hospitality and generous support
during the workshop ``String Theory from a worldsheet perspective,''
where part of this work was done.
This work was supported in part
by a Grant-in-Aid for Scientific Research~(C)~17K05408
from the Japan Society for the Promotion of Science (JSPS).

\end{document}